# Maximum-power quantum-mechanical Carnot engine

Sumiyoshi Abe[1,2,3]

[1] *Department of Physical Engineering, Mie University, Mie 514-8507, Japan*

[2] *Institut Supérieur des Matériaux et Mécaniques Avancés, 44 F. A. Bartholdi, 72000 Le Mans, France*

[3] *Inspire Institute Inc., Alexandria, Virginia 22303, USA*

**Abstract**

In their work [J. Phys. A: Math. Gen. **33,** 4427 (2000)], Bender, Brody, and Meister have shown by employing a two-state model of a particle confined in the one-dimensional infinite potential well that it is possible to construct a quantum-mechanical analog of the Carnot engine through the changes of both the width of the well and the quantum state in a specific manner. Here, a discussion is developed about realizing the maximum power of such an engine, where the width of the well moves at low but finite speed. The efficiency of the engine at the maximum power output is found to be universal independently of any of the parameters contained in the model.

The importance of the Carnot cycle for the foundations of thermodynamics [1,2] cannot be too much emphasized. Its actual implication is however limited due to the assumption of "equilibrium-preserving" processes, making the Carnot engine very slow and thus the power output contextually infinitesimal. There may be at least two possibilities to achieve finite power. One is to introduce irreversibility into the system in such a way that the engine becomes endoreversible [3]. Curzon and Ahlborn [4] (see also Ref. [2] as well as an earlier work [5]) have considered heat fluxes between the working substance and its surrounding heat baths and have afforded the finite duration of time to each process to obtain the well-known expression for the efficiency at the maximum power: $\eta_{CA} = 1 - \sqrt{T_2 / T_1}$, where $T_1$ and $T_2$ are the temperatures of the hot and cold heat baths. Another possibility is to make speed of the piston movement finite. These concepts have extensively been studied in the literature (e.g., [6-22]). Today, they form the bases for endoreversible thermodynamics and finite-time thermodynamics.

In their work [23], Bender, Brody, and Meister have made an intriguing discussion about a quantum-mechanical analog of the Carnot cycle. They have employed a simple two-state model of a single particle confined in a one-dimensional infinite potential well and have devised a reversible cycle by changing the potential width and the quantum state in a specific way. They have found that the work is extractable from such a cycle and the efficiency is given by

$$\eta = 1 - \frac{E_L}{E_H}, \tag{1}$$

where $E_H$ ($E_L$) stands for the expectation value of the system Hamiltonian along the analog of the isothermal process at high (low) "temperature". This interesting similarity between quantum mechanics and thermodynamics has recently been further elaborated in Ref. [24].

In this paper, we discuss the problem of optimizing the power output of the quantum-mechanical Carnot engine considered in Ref. [23]. We show that the efficiency of the engine at the maximum power is

$$\eta^* = 0.573977952..., \tag{2}$$

which is intrinsic and universal in the sense that it does not depend on any of the parameters contained in the model such as the maximum and minimum values of the potential width.

First, we recapitulate the structure of the quantum-mechanical Carnot engine proposed in Ref. [23]. Let $H$ be the Hamiltonian of a particle with mass $m$ confined in the infinite potential well with width $L$. The stationary Schrödinger equation reads $H|u_n\rangle = E_n|u_n\rangle$, where the energy eigenvalues are given by $E_n = n^2\pi^2\hbar^2/(2mL^2)$ ($n = 1, 2, 3, ...$). An arbitrary state, $|\psi\rangle$, can be expanded in terms of the eigenstates as $|\psi\rangle = \sum_n c_n |u_n\rangle$, where the expansion coefficients satisfy the normalization condition: $\sum_n |c_n|^2 = 1$. The expectation value

$$E = \langle \psi | H | \psi \rangle = \sum_n E_n |c_n|^2 \qquad (3)$$

is identified with the analog of the internal energy. Consider a process, along which both $L$ and the expansion coefficients change. Then, the change of $E$ along the process yields the analog of the first law of thermodynamics:

$$\delta'Q = \delta E + \delta'W, \qquad (4)$$

where $\delta'Q$ and $\delta'W$ are the analogs of the changes of the quantity of heat and work, which are given by

$$\delta'Q = \sum_n E_n \cdot \delta |c_n|^2, \qquad (5)$$

$$\delta'W = -\sum_n \delta E_n \cdot |c_n|^2, \qquad (6)$$

respectively. Since we are considering the thermodynamiclike situation, the time scale of the change of the state is much larger than that of the dynamical one, $\sim \hbar/E$. Therefore, the adiabatic scheme can apply, and thus the change of the state $|\psi\rangle$ is dominantly represented by the change of the expansion coefficients [25], justifying Eq. (5). The two-state model considered in Ref. [23] employs only $|u_1\rangle$ and $|u_2\rangle$. The reversible cycle $A \to B \to C \to D \to A$ is constructed as in Fig. 1. (i) During $A \to B$, the state changes from $|u_1\rangle$ to $|u_2\rangle$. In between, it is $a_1(L)|u_1\rangle + a_2(L)|u_2\rangle$ $\left( \left| a_1(L) \right|^2 + \left| a_2(L) \right|^2 = 1 \right)$, and $E = \left[ \pi^2 \hbar^2 / (2mL^2) \right] \left[ 4 - 3 \left| a_1(L) \right|^2 \right]$ is kept unchanged in analogy with the isothermal process in the classical Carnot cycle. Since $a_1(L_A) = 1$ and $a_1(L_B) = 0$, one has $L_B = 2L_A$ and $\left| a_1(L) \right|^2 = \left( 4 - L^2 / L_A^2 \right) / 3$. The force (i.e., the one-dimensional pressure) is $f_{AB}(L) = -\sum_{n=1,2} (\partial E_n / \partial L) \cdot \left| a_n(L) \right|^2$ $= \pi^2 \hbar^2 / (mL_A^2 L)$, and therefore the work is calculated to be $W_{AB} = \int_{L_A}^{L_B = 2L_A} dL\, f_{AB}(L) = \left[ \pi^2 \hbar^2 / \left( mL_A^2 \right) \right] \ln 2$, which is the analog of the amount of heat absorbed by the system from the high-temperature heat bath, $Q_H$. (ii) During $B \to C$, $\delta' Q = 0$ in analogy with the adiabatic process. That is, the system stays in $\left| u_2 \right\rangle$, and the force is given by $f_{BC}(L) = -\partial \left\langle u_2 \right| H \left| u_2 \right\rangle / \partial L = 4\pi^2 \hbar^2 / (mL^3)$. Therefore, the work is $W_{BC} = \int_{L_B = 2L_A}^{L_C} dL\, f_{BC}(L) = (2\pi^2 \hbar^2 / m) \left[ 1 / \left( 4L_A^2 \right) - 1 / L_C^2 \right]$.

(iii) During $C \to D$, the state changes from $\left| u_2 \right\rangle$ to $\left| u_1 \right\rangle$. In between, it is $b_1(L) \left| u_1 \right\rangle$ $+ b_2(L) \left| u_2 \right\rangle$ $\left( \left| b_1(L) \right|^2 + \left| b_2(L) \right|^2 = 1 \right)$, and $E = \left[ \pi^2 \hbar^2 / (2mL^2) \right] \left[ 4 - 3 \left| b_1(L) \right|^2 \right]$ is kept unchanged as in (i). $b_1(L_C) = 0$ and $b_1(L_D) = 1$, and therefore one has $L_D = L_C / 2$ and $\left| b_1(L) \right|^2 = 4 \left( 1 - L^2 / L_C^2 \right) / 3$. The force is given by $f_{CD}(L)$ $= 4\pi^2 \hbar^2 / \left( mL_C^2 L \right)$, and accordingly, the work is $W_{CD} = \int_{L_C}^{L_D = L_C / 2} dL\, f_{CD}(L)$ $= -\left[ 4\pi^2 \hbar^2 / \left( mL_C^2 \right) \right] \ln 2$. (iv) During $D \to A$, $\delta' Q = 0$ as in (ii). So, the system stays in $\left| u_1 \right\rangle$, and the force is $f_{DA}(L) = -\partial \left\langle u_1 \right| H \left| u_1 \right\rangle / \partial L = \pi^2 \hbar^2 / (mL^3)$. The work is $W_{DA} = \int_{L_D = L_C / 2}^{L_A} dL\, f_{DA}(L) = (2\pi^2 \hbar^2 / m) \left[ 1 / L_C^2 - 1 / \left( 4L_A^2 \right) \right] = -W_{BC}$. Thus, during the cycle, $A \to B \to C \to D \to A$, the total amount of work done is $W = (\pi^2 \hbar^2 / m)(1/L_A^2 - 4/L_C^2)\ln 2$, and the efficiency of this engine is

$$\eta = \frac{W}{Q_H} = 1 - 4\left(\frac{L_A}{L_C}\right)^2, \tag{7}$$

which is equal to Eq. (1), where $E_H \equiv \pi^2\hbar^2/(2mL_A^2)$ and $E_L \equiv 4\pi^2\hbar^2/(2mL_C^2)$. The condition

$$r \equiv \frac{L_C}{L_A} \geq 2 \tag{8}$$

should be satisfied in order for the efficiency to be nonnegative.

Now, let us discuss the power output of the quantum-mechanical Carnot engine. Upon realizing the finite power, clearly it is not possible in the present case to introduce heat fluxes into the system because of the absence of heat baths. The one and only possibility here is to make the speed of movement of the well width finite. So, let $v(t)$ and $\tau$ be the speed of the change of $L$ and the total cycle time, respectively. $v(t)$ should be so slow that the adiabatic scheme remains valid. The total amount of movement during a single cycle, $L_{\text{total}}$, is given by

$$L_{\text{total}} = (L_B - L_A) + (L_C - L_B) + (L_C - L_D) + (L_D - L_A)$$

$$= 2(L_C - L_A)$$

$$\equiv \int_0^\tau dt\, v(t) = \bar{v}\tau, \tag{9}$$

where $\bar{v}$ is the average speed. That is, the total cycle time is expressed as follows:

$$\tau = \frac{2}{\bar{v}}(L_C - L_A). \tag{10}$$

Therefore, the power output, $P = W/\tau$, after a single cycle is

$$P = \frac{\pi^2 \hbar^2 \bar{v} \ln 2}{2mL_A^3} \frac{r^2 - 4}{r^3 - r^2}, \tag{11}$$

where $r$ is given in Eq. (8).

Our idea is to maximize the power in Eq. (11) by controlling $r$, with $L_A$ and $\bar{v}$ being fixed. The maximization condition gives rise to the cubic equation

$$r^3 - 12r + 8 = 0, \tag{12}$$

which has the following three real solutions:

$$r_1 = 4\cos\frac{2\pi}{9}, \quad r_2 = 4\cos\frac{4\pi}{9}, \quad r_3 = 4\cos\frac{8\pi}{9}. \tag{13}$$

Taking into account the condition in Eq. (8), among the three, the physical solution is seen to be $r_1$.

Therefore, the efficiency in Eq. (7) at the maximum power output is found to be given by

$$\eta^* = 1 - \frac{1}{4\cos^2(2\pi/9)}, \tag{14}$$

leading to Eq. (2), which is intrinsic/universal in the present model in the sense that it is independent of the system parameters such as the maximum and minimum values of the potential width. This is the main result of the present work.

Comparing the numerical value in Eq. (2) with that in Eq. (1), we immediately see that the value in Eq. (1) is not necessarily larger than that in Eq. (2). In other words, the efficiency of the quantum-mechanical Carnot engine does not have to be maximum: depending on the ratio, $r = L_C / L_A$, $\eta$ in Eq. (7) can be both larger and smaller than $\eta^*$ in Eq. (2). This is due to the fact that the second law is absent in the similarity between quantum mechanics and thermodynamics. The point is deeply related to the concepts of entropy and temperature. The von Neumann entropy, $S^{vN} = -k_B \text{Tr}(\rho \ln \rho)$ (with $k_B$ being Boltzmann's constant), identically vanishes if the density matrix $\rho$ is of a pure state, making it impossible to directly formulate the second law. Instead of the von Neumann entropy, one might consider, e.g., the Shannon entropy, $S^S = -k_B \sum_n |c_n|^2 \ln |c_n|^2$, where $c_n$'s are expansion coefficients, $|\psi\rangle = \sum_n c_n |u_n\rangle$, for example. However, again it is not possible to introduce an analog of temperature through such an entropy. This issue has recently been studied in Ref. [24], where a transmutation of pure-state quantum mechanics into thermodynamics is found to be induced by the Clausius equality.

In conclusion, we have discussed the power output of the two-state quantum-mechanical Carnot engine composed of a particle confined in the one-dimensional infinite potential well. We have calculated the efficiency of the engine at the maximum power and found that it is universal independently of any of the parameters contained in the model.

## ACKNOWLEDGMENT

This work was supported in part by a Grant-in-Aid for Scientific Research from the Japan Society for the Promotion of Science.

———————————————

Figure Caption

FIG. 1 The quantum-mechanical Carnot cycle depicted in the plane of the width ($L$) and force ($f$). The quantum states and the values of the potential width are as follows: $|u_1\rangle$ and $L_A$ at A, $|u_2\rangle$ and $L_B = 2L_A$ at B, $|u_2\rangle$ and $L_C$ at C, and $|u_1\rangle$ and $L_D = L_C/2$ at D.

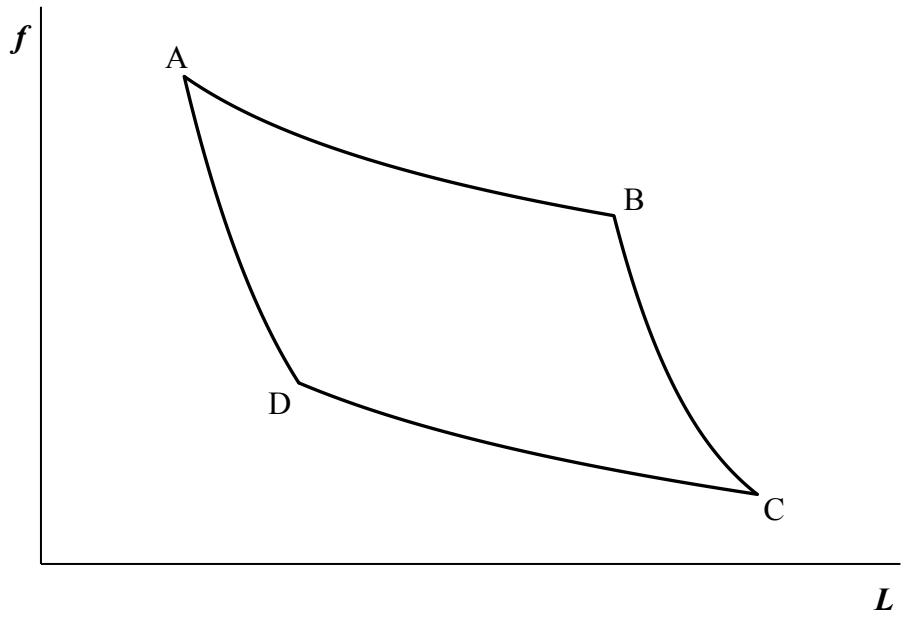

Figure 1